\documentstyle[12pt,epsf]{article}

\newcommand\pubnumber{SLAC-PUB-7681}
\newcommand\pubdate{December, 1997}

\def\Title#1{\begin{center} {\Large #1 } \end{center}}
\def\Author#1{\begin{center}{ \sc #1} \end{center}}
\def\Address#1{\begin{center}{ \it #1} \end{center}}

\def\submit#1{\begin{center}Submitted to {\sl #1} \end{center}}
\def\doeack{\footnote{Work supported by the Department of Energy,
                     contract DE--AC03--76SF00515.}}
\def\SLAC{Stanford Linear Accelerator Center\\
    Stanford University, Stanford, California 94309 USA}
\newcommand\pubblock{\rightline{\begin{tabular}{l} \pubnumber\\
         \pubdate  \end{tabular}}}
\newenvironment{Abstract}{\begin{quotation} \begin{center}
                       ABSTRACT
     \end{center}\bigskip  }{\end{quotation}}
\def\beq{\begin{equation}}
\def\eeq#1{\label{#1}\end{equation}}
\def\eeqn{\end{equation}}
\def\beqa{\begin{eqnarray}}
\def\eeqa#1{\label{#1}\end{eqnarray}}
\def\eeqan{\end{eqnarray}}

\def\Acknowledgements{\bigskip  \bigskip \begin{center} \begin{large}
             \bf ACKNOWLEDGEMENTS \end{large}\end{center}}
\begin{document}
\begin{titlepage}
\pubblock

\vfill
\Title{Transmission of Supersymmetry Breaking from a 4-Dimensional
            Boundary}
\vfill
\Author{Eugene A. Mirabelli and Michael E. Peskin\doeack}
\Address{\SLAC}
\vfill
\begin{Abstract}
In the strong-coupling limit of the heterotic string theory constructed
by Ho\v{r}ava and Witten, an 11-dimensional supergravity theory is coupled to 
matter multiplets confined to 10-dimensional mirror planes. 
 This structure 
suggests that realistic unification models are obtained, after 
compactification
of 6 dimensions, as theories of 5-dimensional supergravity in an interval,
coupling to matter fields on 4-dimensional walls.  Supersymmetry 
breaking may be communicated from one boundary to another by the 5-dimensional
fields.  In this paper, we study a toy model of this communication in 
which 5-dimensional super-Yang-Mills theory in the bulk couples to 
chiral multiplets on the walls.  Using the auxiliary fields of the Yang-Mills
multiplet, we find a simple algorithm for coupling the bulk and boundary
fields. We demonstrate two different mechanisms for generating soft 
supersymmetry breaking terms in the boundary theory.  We also compute the 
Casimir energy generated by supersymmetry breaking.
\end{Abstract}
\medskip
\submit{Physical Review {\bf D}}

\vfill
\end{titlepage}
\def\thefootnote{\fnsymbol{footnote}}
\setcounter{footnote}{0}
\section{Introduction}

In their recent investigation of the structure of strongly-coupled heterotic
string theory, Ho\v{r}ava and Witten have introduced a new paradigm for 
models of unification \cite{hw1,hw2,unification}.
  To construct the strong-coupling limit of the 
heterotic string, they began from the 11-dimensional strong-coupling limit
of the Type IIA string theory. They compactified this model on $S^1/Z_2$,
that is, on an interval of length $\ell$ bounded by mirror (orientifold)
 planes.   They then argued that a 10-dimensional $E_8$ super-Yang-Mills 
theory appears on each plane.  The final structure is a set of two 
$E_8$ gauge theories on the mirror planes, coupling to supergravity in the 
interior of the interval. 

This arrangement had an immediate phenomenological
advantage over the weakly coupled $E_8\times E_8$ heterotic string 
theory \cite{unification}.
 When $\ell$ was increased, the low-energy value of Newton's 
constant decreased proportional to $1/\ell$, while the $E_8$ gauge coupling
remained fixed.  Thus, by adjustment of $\ell$, one could arrange a 
unification of gauge and gravitational couplings.  Choosing a 
 large value of $\ell$ relative to the 11-dimensional Planck scale
justified the use of only field-theoretic, and not intrinsically
string-theoretic, degrees of freedom.  At the same time, Ho\v{r}ava and Witten
obtained reasonable values for the gauge and gravitational couplings
for values of $\ell$ not so large, in the sense that
 both of these scales could be considered to be 
of the order of the grand unification scale of
$2\times 10^{16}$ GeV inferred from the values of the gauge couplings at
low energy. 

In a realistic phenomenology, 6 of the transverse 10 dimensions should be
compactified.  Then one would obtain a 5-dimensional theory on an interval
with mirror-plane boundaries.  Plausibly, this theory could be described
as a  5-dimensional supergravity field theory, perhaps with some additional
bulk supermultiplets, coupling to matter supermultiplet fields on the walls.
If $\ell$ is the largest dimension in this geometry, it is reasonable that
the theory should make sense in the limit in which $\ell$ is taken to be 
nonzero while the finite size of the 6-dimensional compactification space is 
ignored.

Ho\v{r}ava and Witten introduced another very interesting idea on the nature
of these compactifications.  They pointed out that the matter theory could
be at strong coupling on one boundary, and could break supersymmetry 
spontaneously there.  Then the supersymmetry-breaking effects could be
communicated to the other boundary by 11- or 5-dimensional fields.  In this
way, the theory on one boundary would become the `hidden sector' for the
phenomenological supersymmetry theory on the other boundary.

Ho\v{r}ava tried
to make this mechanism of communication 
 explicit by exhibiting a term in the 
11-dimensional Lagrangian which coupled the gaugino condensate on the
boundary to the 3-form gauge field $C_{ABC}$ of the bulk supergravity
theory \cite{allelse}.  He found that this term had a perfect-square structure
\begin{equation}
   \Delta L =  -{1\over 12 \kappa^2}\int d^{11} x \left( \partial_{11} 
         C_{ABC} - {\sqrt{2}\over 16\pi}\left({\kappa\over 4\pi}\right)^{2/3}
                       \overline{\chi} \Gamma_{ABC} \chi\ \delta(x^{11}) \right)^2 \ ,
\label{Hsform}\end{equation}
where $\chi$ is the 10-dimensional gaugino and $8\pi\kappa^2$ is the 
11-dimensional Newton constant. Ho\v{r}ava argued that, if the gaugino
bilinear obtains a nonzero value, there is no solution for $C_{ABC}$
which is consistent with supersymmetry.

Ho\v{r}ava's observations raise two  interesting questions of principle.
The first concerns the structure of (\ref{Hsform}).  We might want to 
know how the delta function on the boundary shown in (\ref{Hsform}) arises.
The square of this term integrated over $x^{11}$ gives
a factor $\delta(0)$ in the boundary Lagrangian.  It is a puzzling issue
whether this term is reasonably included in a purely field-theoretic
description of the Ho\v{r}ava-Witten compactification, or whether the
presence of this term
 implies that any such field-theoretic description is incomplete.

The second question
comes from the fact that the communication between the two 
boundaries comes from the 3-form gauge field, a rather exotic agent.
From the general form  of the potential energy in supergravity,
 the 4-dimensional theory which we would obtain by compactifying 6 dimensions
and then taking the limit $\ell \to 0$ must contain a direct coupling of
the superpotentials on the two boundaries.  We would like to know how this
coupling arises, and how much of this coupling is present in the
compactified theory before we take the limit $\ell\to 0$.  In the standard
approach to supersymmetry breaking mediated by supergravity, this coupling
is the source of the soft supersymmetry-breaking mass terms for squarks
and sleptons.  It would be wonderful if the presence of an extended 
fifth dimension had specific consequences for the superparticle mass 
spectrum which could be verified experimentally.

We have tried to find the answers to these questions
by studying a toy model in which supergravity is replaced by a 
Yang-Mills supermultiplet. Consider, then, 5-dimensional super-Yang-Mills
theory on an interval of length $\ell$ bounded by mirror planes, 
coupled to chiral multiplets 
$\phi$, $\phi'$ on the 4-dimensional boundaries.  In the limit $\ell\to 0$,
this theory must have a potential energy with the $D$-term 
contribution
\begin{equation}
         V = {g^2\over 2} (Q\phi^\dagger \phi +
  Q' \phi^{\prime\dagger} \phi^\prime)^2 \ ,
\label{Dterm}\end{equation}
where $g$ is the effective 4-dimensional coupling constant and $Q$, $Q'$
are the gauge charges of $\phi$, $\phi'$.
So we can ask in this system also how much of the coupling between boundaries
which is required in the limit $\ell\to 0$ survives when $\ell$ is 
kept nonzero.   The related problem of coupling a 5-dimensional
of hypermultiplets to a superpotential on the boundary has been studied
previously by Sharpe \cite{sharpe}, but, we feel, without giving 
the insight that we are seeking.

A convenient strategy for coupling  5-dimensional supermultiplets to a 
4-dimensional boundary is to work with the off-shell supermultiplets, 
including auxiliary fields.  Under straightforward dimensional 
reduction, 5-dimensional multiplets reduce to 4-dimensional $N=2$
supermultiplets.  A mirror plane, or orientifold, 
 obtained by a $Z_2$ identification has
lower supersymmetry, and so on such a plane a 5-dimensional 
multiplet should reduce to a 4-dimensional $N=1$ supermultiplet.
Nevertheless, if we have the correct off-shell multiplet, we can couple
it straightforwardly to 4-dimensional fields on the boundary.  

In Section 2, we will present the necessary formalism for coupling a
5-dimensional super-Yang-Mills multiplet to an orientifold boundary.
We  will identify the off-shell 4-dimensional supermultiplet
which couples to the boundary fields and use this multiplet to construct the
4-dimensional boundary Lagrangian.
In Section 3, we will discuss the role of the $\delta(0)$ terms which
appear in this Lagrangian, illustrating our conclusions by some 
explicit one- and two-loop calculations.  

In Section 4, we will use the formalism that we have developed to 
discuss the communication of supersymmetry breaking from one boundary
to the other.  We will first analyze the case in which supersymmetry
is spontaneously  broken by a Fayet-Iliopoulos term on one boundary.
Then we will present a more involved example in which supersymmetry
is communicated by loop diagrams which span the fifth dimension.

If supersymmetry is spontaneously broken, the vacuum energy can be 
nonzero.  In general, the vacuum energy will contain a term,
called the Casimir energy \cite{Casimir}, which depends
on the separation of the two 
boundaries.  
  In the eventual application to supergravity, this dependence
is needed to fix the size of the compact geometry.  Though the case of 
5-dimensional Yang-Mills theory is simpler than that of supergravity
in several respects, it is still interesting to compute the 
Casimir energy for this case.  In Section 5, we evalute this energy
for the models of the communication of supersymmetry-breaking studied
in Section 4 and note the similarities of the two computations.

In Sections 3 through 5, we will be carrying out weak-coupling 
perturbation theory computations in the nonrenormalizable 5-dimensional
Yang-Mills theory.  Our attitude toward this nonrenormalizability is a 
pragmatic one; we will be pleased if quantities of physical interest
turn out to be ultraviolet-finite at the leading order of perturbation
theory.
That will be true in our explicit calculations of the scalar mass term
and the Casimir energy.  Presumably, the higher-order corrections to these
computations are cutoff-dependent and are regulated by the underlying 
string theory.  In this paper, we will not be concerned with effects
beyond the leading nontrivial order.

Finally, in Section 6, we will discuss the relation of our formalism 
to Ho\v{r}ava's analysis and give an explanation of the coupling shown in 
(\ref{Hsform}).

Our approach to the Ho\v{r}ava-Witten theory complements the many attempts to 
understand the structure of this theory by direct analysis of the
11-dimensional Lagrangian.  General properties of the strong-coupling
limit of the heterotic string theory have been discussed in 
\cite{Banks,ABF,Choi}.  More explicit studies of the compactification of the 
Ho\v{r}ava-Witten theory have been discussed recently 
by several groups.  Some of these analysis \cite{LLN,NOY,LT,ChoiandM,Ovrut}
have emphasized the 
connection to the venerable  mechanism of supersymmetry breaking in string 
theory by $E_8$ gaugino condensation \cite{DSRW},
 while others \cite{DG,IQ} have relied
on the Scherk-Schwarz mechanism \cite{SS} in the bulk 
to provide a new source of 
supersymmetry breaking.  Brax and Turok \cite{BT}
have contributed an observation 
on the possibility of large hierarchies in the 5-dimensional geometry, if
all of the relevant 5-dimensional fields can  be made massive. 
 We hope that the methods of analysis that we
introduce here, when generalized to supergravity, will clarify the many
possible sources of supersymmetry breaking which may contribute in  the 
Ho\v{r}ava-Witten approach to unification.

\section{Bulk and boundary supermultiplets}
\label{sec-transforms}

In this section, we will set up the formalism for coupling 5-dimensional 
super-Yang-Mills theory to an orientifold boundary.  The 5-dimensional
Yang-Mills multiplet contains a vector field $A^M$, a real scalar field
$\Phi$, a gaugino $\lambda^i$. 

 In this paper, capitalized indices
$M,N$ run over 0,1,2,3,5, lower-case indices $m$ run over 0,1,2,3, and 
$i$, $a$ are internal $SU(2)$ spinor and vector indices,
 with $i = 1,2$, $a= 1,2,3$.  We use a timelike metric $\eta_{MN}$ = 
diag(1,-1,-1,-1,-1) and take the following basis for the 
Dirac matrices:
\begin{equation}
   \gamma^M = \left( \pmatrix{0 & \sigma^m \cr \overline{ \sigma}^m & 0\cr}\ , 
\ \pmatrix{ -i & 0 \cr 0 & i} \right) \ ,
\label{gammas}\end{equation}
where $\sigma^m = (1, \vec \sigma)$, $\overline{ \sigma}^m = (1, -\vec \sigma)$.
 Though it is conventional in the literature to use
raised and lowered spinor indices, we find it less confusing to write out 
explicitly the $2\times 2$ and $4\times 4$ charge conjugation matrices
\begin{equation}
        c = -i\sigma^2 \ , \qquad   C = \pmatrix{ c & 0 \cr 0 & c \cr}\ .
\label{cdef}\end{equation}

In 5-dimensional supersymmetry, it is convenient to rewrite 4-component
Dirac spinors as symplectic-Majorana spinors, Dirac fermions which carry an
$SU(2)$ spinor index and satisfy the constraints
\begin{equation}
             \psi^i = c^{ij} C \overline{ \psi}^{jT} \ . 
\label{symMaj}\end{equation}
A symplectic-Majorana spinor can be decomposed into 4-dimensional chiral
spinors according to 
\begin{equation}
        \psi^i = \pmatrix{ \phi_L^i \cr \phi^i_R } 
\label{sMdecomp}\end{equation}
where $\phi_{L,R}^i$ are two-component spinors connected by 
\begin{equation}
      \phi_L^i = c^{ij} c \phi^{j*}_R  \ , \qquad 
      \phi_R^i = c^{ij} c \phi^{j*}_L \ .
\label{chirels}\end{equation} 
Symplectic-Majorana spinors $\psi^i$, $\chi^i$ satisfy the identity
\begin{equation}
   \overline{ \psi}^i \gamma^M \cdots \gamma^P \chi^j  = 
        - c^{ik} c^{j\ell}     
 \overline{ \chi}^\ell \gamma^P \cdots \gamma^M \psi^k  \ , 
\label{SMid}\end{equation}
including the minus sign from fermion interchange.

In this notation, the 5-dimensional Yang-Mills multiplet is extended to an
off-shell multiplet by adding an $SU(2)$ triplet $X^a$ of real-vauled
auxiliary fields \cite{basicsusy}.  Write the members of the multiplet as
matrices in the adjoint representation of the gauge group: $A^M = A^{MA} t^A$,
{\it etc}.  The the supersymmetry 
transformation laws are
\begin{eqnarray}
{\delta_{\xi}} A^M       & = & i {\overline{\xi}}^i \gamma^M \lambda^i  \nonumber \\ 
{\delta_{\xi}} \Phi      & = & i {\overline{\xi}}^i \lambda^i  \nonumber \\ 
{\delta_{\xi}} \lambda^i & = & ( \sigma^{MN} F_{MN} 
	- \gamma^M D_M \Phi ) \xi^i 
		- i( X^a \sigma^a )^{ij} \xi^j \nonumber \\ 
{\delta_{\xi}} X^{a}     & = & {\overline{\xi}}^i (\sigma^a)^{ij} 
		\gamma^M D_M \lambda^j - i [ \Phi , {\overline{\xi}}^i 
                       (\sigma^a)^{ij} \lambda ^j ]\ ,
\label{fivedtrans}\end{eqnarray}
where the symplectic-Majorana spinor $\xi^i$ is the supersymmetry parameter,
$ D_M \Phi \equiv \partial_M \Phi - i[ A_M, \Phi ] $ (and similarly for
 $ D_M \lambda $ ), and 
$ \sigma^{MN} \equiv \frac{1}{4} [\gamma^M , \gamma^N] $.

Now we would like to project this structure down to a 4-dimensional $N=1$
supersymmetry transformation acting on fields on the orientifold wall.
In a field theory description, an orientifold at $x^5 = 0$ is described
by imposing the boundary condition on bulk fields $a(x,x^5)$
\begin{equation} 
             a(x^m,x^5) = P\, a(x^m,-x^5)
\label{parityfive}\end{equation}
where $P$ is an intrinsic parity equal to $\pm 1$.  The quantum number $P$
must be assigned to fields in such a way that it leaves the bulk Lagrangian
invariant.  Then fields of $P= -1$ vanish on the walls but have nonvanishing
derivatives $\partial_5 a$.  Note that, since $A^5$ vanishes on the boundary,
$\partial_5 = D_5$ on the boundary and $\partial_5 a$ is gauge-covariant.

Let $\xi^1_L$ be the supersymmetry parameter of the $N=1$ supersymmetry
transformation on the wall.  Then the $P$ assignments of the fields in the bulk
supermultiplet are determined as follows:
\begin{equation}
\begin{tabular}{|c|c|c|}                     
 & \ $P=  +1$ \   & \ $P =  -1$ \ \\ \hline\hline
$\xi$       & $\xi^1_L$     & $\xi^2_L$     \\ \hline
$A^M$       & $A^m$         & $A^5$         \\ \hline
$\Phi$      & -             & $\Phi$        \\ \hline
$\lambda^i$ & $\lambda^1_L$ & $\lambda^2_L$ \\ \hline
$X^a$       & $X^3$         & $X^{1,2}$     \\ \hline
\end{tabular}
\label{ptable}\end{equation}
On the wall at $x^5 = 0$, the supersymmetry transformation 
(\ref{fivedtrans}) reduces to the following transformation of the
even-parity states generated by $\xi^1_L$:
\begin{eqnarray}
{\delta_{\xi}} A^{m}       & = 
	& i \xi^{1\dagger}_L \overline{\sigma}^m \lambda^1_L 
			- i   \lambda^{1\dagger}_L \overline{\sigma}^m \xi^1_L 
			\nonumber \\ 
{\delta_{\xi}} \lambda^1_L & = 
	& \sigma^{mn}F_{mn}\xi^1_L -i (X^3 - \partial_5 \Phi) 
			\xi^1_L \nonumber \\ 
{\delta_{\xi}} X^{3}       & = 
	&  \xi^{1\dagger}_L \overline{\sigma}^m D_m \lambda^1_L
			-i \xi^{1\dagger}_L c \partial_5 \lambda^{2*}_L
			+ h.c. \nonumber \\ 
{\delta_{\xi}} \partial_5 \Phi    & = 
	& -i \xi_L^{1T} c\partial_5 \lambda_L^2 - i
 \xi_L^{1 \dagger} c \partial_5 \lambda_L^{2*} \ .
\label{fourdtrans}\end{eqnarray}
The last two equations imply
\begin{equation}
{\delta_{\xi}} (X^{3}-\partial_5\Phi) 
	=  \xi^{1\dagger}_L \overline{\sigma}^m D_m \lambda^1_L
			+ h.c. \ \ .
\label{Dtrans}\end{equation}
These are just the transformation laws for an $N=1$ 4-dimensional
 vector multiplet \cite{wessbagger}, with $A^m$, $\lambda^1_L$, 
and $(X^3 - \partial_5\Phi)$ transforming as the vector, gaugino, and 
auxiliary D fields.

The appearance of the quantity $\partial_5\Phi$ in 
the auxiliary field should not
be a surprise.  It is the expectation value of this quantity that breaks
supersymmetry in Scherk-Schwarz mechanism \cite{SS}.  Thus, $\partial_5\Phi$
should appear in the order parameter of supersymmetry breaking.

Now it is obvious how to couple the 5-dimensional gauge multiplet to 
4-dimensional chiral multiplets on the boundary.  We write the Lagrangian
as 
\begin{equation} 
  S =   \int d^5x \left\{ {\cal L}_5 
	+ \sum_i \delta(x^5 - x^5_i){\cal L}_{4i} \right\}\ ,
\label{totalS}\end{equation}
where the sum includes the  walls at $x^5_i = 0, \ell$. The bulk 
Lagrangian should be  the standard one for a 5-dimensional super-Yang-Mills 
multiplet,
\begin{eqnarray}
{\cal L}_{5}   & = 
	&{1\over g^2}\biggl( -\frac{1}{2}{\mbox{\rm tr}}(F_{MN})^2 
	+ {\mbox{\rm tr}}(D_M \Phi)^2
   + {\mbox{\rm tr}}(\overline{\lambda} i  \gamma^M D_M \lambda) \nonumber \\ 
	&   & 
  + \ {\mbox{\rm tr}}(X^a)^2 - {\mbox{\rm tr}}
	(\overline{ \lambda} [\Phi,\lambda])\biggr) \ ,
\label{fivedL}\end{eqnarray}
with ${\mbox{\rm tr}} [t^A t^B ] = \delta^{AB}/2$.
The bulk fields should be constrained to satisfy the boundary conditions
(\ref{parityfive}) at  the walls.   Since the 
supersymmetry generated by $\xi_L^1$ relates fields with the same
boundary conditions, this $N=1$ supersymmetry is an invariance of 
(\ref{fivedL}) under the constraint.

The boundary Lagrangian should have the standard form of a four-dimensional
chiral model built from supermultiplets $(\phi,\psi_L,F)$.   Here 
and in the rest of the paper, it is important to distinguish boundary
chiral scalar fields, which we will label by $\phi$, from the bulk
scalar field $\Phi$.  The explicit form of this boundary Lagrangian is
\begin{eqnarray}
{\cal L}_4 & = 
	& D_m\phi^\dagger D^m\phi + \psi_L^\dagger i 
\overline{\sigma}^m D_m \psi_L
              + F^\dagger F \nonumber \\ 
    & &  - \sqrt{2}i\left( \phi^\dagger \lambda_L^T c \psi_L 
              + \psi^\dagger c \lambda^*_L \phi\right) + \phi^\dagger D \phi
                      \ ,
\label{fourdL}\end{eqnarray}
with $D_m = (\partial_m - iA_m)$, and with 
with the gauge fields $(A_m,\lambda_L,D)$ replaced by the boundary values
of the bulk fields $(A_m, \lambda^1_L, X^3 - \partial_5 \Phi)$. 
 The boundary
Lagrangian ${\cal L}_4$ is invariant by itself under an $N=1$ supersymmetry
transformation of the boundary fields and the supersymmetry transformation
(\ref{fourdtrans}) of the bulk fields.  Thus, the complete action
(\ref{totalS}) is $N=1$ supersymmetric.

In principle, we could add to (\ref{totalS}) additional terms involving
a four-dimensional integral of the boundary values of the vector fields, or
terms coupling the chiral fields to 
higher $\partial_5$ derivatives of the vector fields at the boundary.  These 
terms would correspond to contributions that are more singular at the boundary
that the ones we have considered. For our present purposes, we only 
 point out that these terms are not necessary for supersymmetry, and we 
neglect them from here on.  We will show in explicit calculations that the
terms we have written suffice to give amplitudes which converge to the correct
4-dimensional limits as $\ell \to 0$.

With the action (\ref{totalS}),
the boundary scalar field $\phi$ at $x^5 = 0$ 
couples to the auxiliary field $X^3$
through the terms
\begin{equation}
  \int d^5x \left\{ {1\over g^2}{\mbox{\rm tr}} 
	(X^3)^2  + \delta(x^5) \phi^\dagger
   (X^3 - \partial_5 \Phi) \phi \right\} \ .
\label{Xphi}\end{equation}
The field $\Phi$ is a dynamical field in the interior, but $X^3$ is an 
auxiliary field and may be integrated out.  This gives a boundary 
Lagrangian of the form
\begin{equation}
   \int d^4x \left\{ - \phi^\dagger (\partial_5\Phi) \phi 
	- \frac{1}{2} (\phi^\dagger t^A
       \phi)^2 \delta(0) \right\} \ .
\label{dofz}\end{equation}

Thus, our formalism does contain singular terms proportional to $\delta(0)$
on the boundary, which arise naturally from integrating out the auxiliary
fields.  In principle, the complete description of the orientifold wall
in string theory could contain additional couplings involving higher
derivatives $\partial_5$ of the bulk fields and representing a finite thickness
and a shape of the wall.  However, the Lagrangian we have written, with 
the $\delta(0)$ but no additional  singular terms, is a completely
self-consistent supersymmetric system.

\section{Bulk and boundary perturbation theory}
\label{sec-calculations}

In the previous section, we have found that singular terms proportional 
to $\delta(0)$ on the boundary arise naturally when bulk and boundary
fields are coupled supersymmetrically.  What is still unclear is whether
these terms can lead to sensible results when one performs computations 
in this theory, or whether these terms signal the breakdown of a purely
field-theoretic description.  We believe that these singular terms do make
sense at the field theory level.  Their role is to provide counterterms 
which are necessary in explicit calculations to maintain supersymmetry.
In this section, we will illustrate this conclusion with some explicit
calculations in perturbation theory.

%%%%%%%%%%%%%%%%%%%%%%%%%%%%%%%%%%%%%%%%%%%%%%%%%%%%%%%%%%%%%%%%%%%%%%
\begin{figure}
\begin{center}
\leavevmode
{\epsfxsize=4.00truein \epsfbox{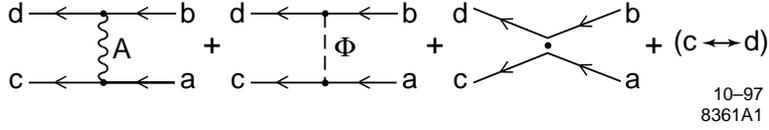}}
\end{center}
 \caption{Feynman diagrams contributing to the scattering process
      $\phi\phi\to \phi\phi$.}
\label{fig:phiphi}
\end{figure}
%%%%%%%%%%%%%%%%%%%%%%%%%%%%%%%%%%%%%%%%%%%%%%%%%%%%%%%%%%%%%%%%%%%%%%

As a first, simplest, example, consider the scattering amplitude for 
scalars on a wall.  The Feynman diagrams contributing to the process
$\phi_a + \phi_b \to \phi_c + \phi_d$ are shown in Figure \ref{fig:phiphi}.
The propagator of a free massless bulk field is
\begin{equation}
  \left\langle{ a(x,x^5) a(y,y^5)} \right\rangle = 
	\int_{k5} {i \over k^2 - (k^5)^2}\,
 e^{-ik\cdot (x-y)} (e^{ik^5(x^5-y^5)}
     + P e^{ik^5(x^5+y^5)}) \ ,
\label{fiveprop}\end{equation}
where 
\begin{equation}
   \int_{k5} = \int {d^4k\over (2\pi)^4} {1\over 2\ell}\sum_{k^5} \ ,
\label{intdef}\end{equation}
with $k^5$ summed over the values $ \pi m /\ell$, $m =$ integer.
Here and in the rest of our discussion, $k$ represents a the 4-dimensional
momentum components of $k^M$.

The sum of diagrams in Figure \ref{fig:phiphi} is given by
\begin{eqnarray}
i{\cal M}(\phi_a + \phi_b \to \phi_c + \phi_d) &= &
  -ig^2  t^A_{ca} t^A_{db} \biggl( {1\over 2\ell}\sum_{k^5} {(k^5)^2\over
                     (p_c - p_a)^2 - (k^5)^2}
         + \delta(0) \nonumber \\    & & 
 -  {1\over 2\ell}\sum_{k^5} {(p_c + p_a)\cdot (p_d +  p_b)\over 
   (p_c - p_a)^2 - (k^5)^2}\biggr)
      + (c \leftrightarrow d) \ . 
\label{twototwo}\end{eqnarray}
If we represent
\begin{equation}
    \delta(0) = {1\over 2\ell}\sum_{k^5} 1 = 
 {1\over 2\ell}\sum_{k^5} { k^2 - (k^5)^2\over k^2 - (k^5)^2}\ ,
\label{repdelta}\end{equation}
the first two terms have a neat cancellation and we find the finite 
result
\begin{eqnarray}
i{\cal M}(\phi_a + \phi_b \to \phi_c + \phi_d) & = &  
  -ig^2  t^A_{ca} t^A_{db} \left( {1\over 2\ell} \sum_{k^5}
 {(p_c - p_a)^2 - (p_c + p_a) \cdot (p_d + p_b)\over 
   (p_c - p_a)^2 - (k^5)^2} \right) + (c \leftrightarrow d) \nonumber \\ 
 & = & 
  -ig^2  t^A_{ca} t^A_{db} \left( {1\over 2\ell} \sum_{k^5}
 {-2u\over 
   t - (k^5)^2} \right)+ (c \leftrightarrow d) \ . 
\label{twototwotwo}\end{eqnarray}
If $\ell \to 0$ with the dimensionless coupling $g^2/\ell$ fixed,
the terms with $k^5 \neq 0$ become negligible.  Then we recover the 
scalar particle scattering amplitude of a 4-dimensional $N=1$ supersymmetric
gauge theory.

%%%%%%%%%%%%%%%%%%%%%%%%%%%%%%%%%%%%%%%%%%%%%%%%%%%%%%%%%%%%%%%%%%%%%%
\begin{figure}
\begin{center}
\leavevmode
{\epsfxsize=4.00truein \epsfbox{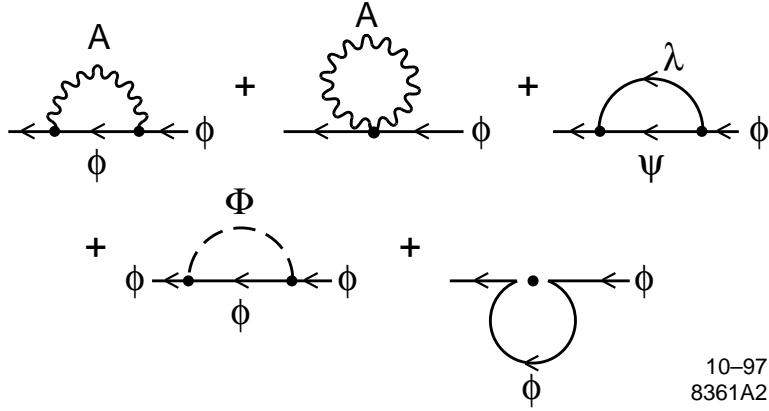}}
\end{center}
 \caption{Feynman diagrams contributing to the $\phi$ self-energy
      at one-loop order.}
\label{fig:phiself}
\end{figure}
%%%%%%%%%%%%%%%%%%%%%%%%%%%%%%%%%%%%%%%%%%%%%%%%%%%%%%%%%%%%%%%%%%%%%%

As a second example, consider the self-energy of the scalar $\phi$, 
computed at the one-loop level.   By supersymmetry, the $\phi$ cannot
obtain a mass in perturbation theory, but it is interesting to see
explicitly how the cancellation occurs.
 The Feynman diagrams for the $\phi$ self-energy 
are shown in Figure \ref{fig:phiself}.  The first four diagrams
all involve one  field that propagates in four dimensions and one 
field that propagates 
 in the fifth dimension.  Thus, if $p$ is the external 4-momentum, all of 
these diagrams will have the structure 
\begin{equation} 
  -i M^2(p^2) = 
  g^2 t^A t^A \int_{k5} {1\over k^2 - (k^5)^2} {1\over (p-k)^2} 
 N(k,k^5,p)\   ,
\label{twopointform}\end{equation}
where $N$ is a polynomial in momenta.  Using the representation 
(\ref{repdelta}), we can bring the last diagram into this form as well.
Then the five diagrams give contributions 
\begin{eqnarray}
   N &= &
 -(2p-k)^2 + 4(p-k)^2 - 4k\cdot(k-p) + (k^5)^2 + (k^2 - (k^5)^2)\nonumber \\ 
     &= & 0 \ .
\label{Ncancel}\end{eqnarray}
Here the $\delta(0)$ term enters quite explicitly as a counterterm which 
cancels the singluar behavior of the $\Phi$ exchange diagram and thus    
allows the complete cancellation required by supersymmetry.

%%%%%%%%%%%%%%%%%%%%%%%%%%%%%%%%%%%%%%%%%%%%%%%%%%%%%%%%%%%%%%%%%%%%%%
\begin{figure}
\begin{center}
\leavevmode
{\epsfxsize=4.00truein \epsfbox{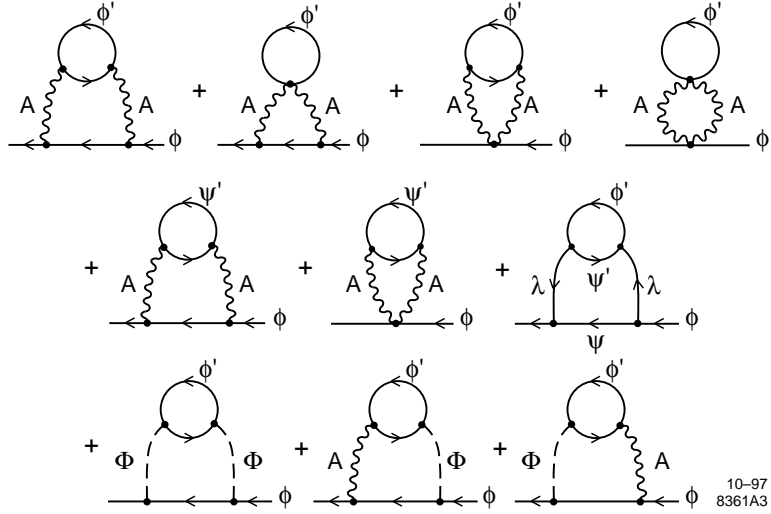}}
\end{center}
 \caption{Feynman diagrams contributing to the mass shift of a scalar
 $\phi$ on one wall due to loop effects of the supermultiplet on the
        other wall.}
\label{fig:twoloop}
\end{figure}
%%%%%%%%%%%%%%%%%%%%%%%%%%%%%%%%%%%%%%%%%%%%%%%%%%%%%%%%%%%%%%%%%%%%%%

To prepare for the next section, it will be useful to illustrate one more
cancellation required by supersymmetry. Consider the renormalization of the
mass of a scalar $\phi$ on one wall due to loop diagrams involving the 
supermultiplet on the other wall.  This mass shift is given by the 
sum of the two-loop diagrams shown in Figure~\ref{fig:twoloop}. In our 
discussion of these diagrams, we will assign the chiral fields $\phi$ at
$x^5 = 0$ to the representation $R$ of the gauge group, and we will assign
the fields $\phi'$ at $x^5 = \ell$ to the representation $R'$.

The diagrams of Figure~\ref{fig:twoloop}
form a gauge-invariant set. We might characterize them as the diagrams of 
order $g^4 N$, where $N$ is the number of matter multiplets on the second
wall.  Thus, by supersymmetry, their sum must vanish.

It is not difficult to see this explicitly. Since we are interested in 
the shift of the $\phi$ mass term, we can set the external momentum 
equal to zero.  Then let $q$ be the loop momentum of the matter loop on the 
wall at $x^5 = \ell$,  Let $(k,k^5)$ and $(k,{\hat k}^5)$ be the 
momenta carried by the two propagators of the gauge multiplet that 
connect the two walls.  These momenta are quantized, with 
\begin{equation}
         k^5 = \pi n/\ell\ ,\qquad {\hat k}^5 = \pi \hat n/\ell\ , 
\label{nhatdefin}\end{equation}
but  the integers $n$ and $\hat n$ are summed over independently,
 since $k^5$ is not conserved in the interaction of bulk 
fields with the walls.  Then all of the diagrams shown in
 Figure~\ref{fig:twoloop} can be written in the form 
\begin{equation}
  -i M^2 = 
 i g^4 C_2(R)  C(R') \int_q \int_{k55}{ N(k,k^5,{\hat k}^5,q)\over 
 (k^2)(k^2 - (k^5)^2)(k^2 -( {\hat k}^5)^2) (q^2)((q-k)^2)} \ ,
\label{twoloopform}\end{equation}
where $N$ is a polynomial in momenta, $C(R')\delta^{AB} 
= {\mbox{\rm tr}}_{R'} [t^A t^B]$
is the sum 
over the gauge quantum numbers of the multiplet at $x^5=\ell$, the integral
over $q$ is a simple 4-dimensional momentum integral, and 
\begin{equation}
   \int_{k55} = \int {d^4k\over (2\pi)^4} {1\over 2\ell}\sum_{n} \,
 {1\over 2\ell}\sum_{{\hat n}}  \  4 \cdot (-1)^{n + \hat n}  \ .
\label{inttwodef}\end{equation}
This expression includes the $k^5$-dependence of the propagators, obtained
by evaluating (\ref{fiveprop}) at $x^5 = 0$, $y^5 = \ell$.

To see that the diagrams of this set must cancel, it is easiest to 
compare this calculation to the corresponding two-loop mass renormalization
in four dimensions.
The diagrams on the first two lines of  Figure~\ref{fig:twoloop} contain,
from the five-dimensional gauge multiplet,
only the propagators of the 
fields $A_m$ and $\lambda^1_L$
which appear in a 4-dimensional gauge multiplet.  Thus, their
contributions to the numerator polynomial $N$ are exactly those of the 
corresponding diagrams in 4~dimensions. To treat the last three diagrams,
we note the identity
\begin{equation} 
   0 = 
   {1\over 2\ell}\sum_{k^5} (-1)^n =  
 {1\over 2\ell}\sum_{k^5} (-1)^n { k^2 - (k^5)^2\over k^2 - (k^5)^2}\ ,
\label{repdeltazero}\end{equation}
since the second term is a representation of $\delta(x^5) $ evaluated
at $x^5 = \ell$.  Each $\Phi$ propagator comes with a factor $(k^5)^2$, 
due to the couplings (\ref{dofz}) at each wall.  The identity 
(\ref{repdeltazero}) allows us to replace this $(k^5)^2$ by $k^2$.
Then each diagram gives the same contribution to the numerator as the 
corresponding 4-dimensional diagram with the $\Phi$ replaced by a 
$D$-term interaction.  Thus, the numerator polynomial $N$
turns out to be exactly the one that appears in the 4-dimensional
calculation.  

At this point, we know that the integral (\ref{twoloopform}) must vanish.
It is not difficult to evaluate the various contributions to the 
numerator and to see that they cancel. In the Appendix, we give
a formula for the numerator factor
$N$ from which this can be verified explicitly.

\section{Wall to Wall Supersymmetry Breaking} 
\label{sec:walltowall}

We have now described and tested an explicit form for the coupling of 
4-dimensional supermultiplets on the boundary to gauge supermultiplets
in the bulk. Now we can use this formalism to see 
how supersymmetry breaking on one wall is communicated to the other
wall to provide soft supersymmetry-breaking terms.  In this section, we
will give two examples of such communication, one through a direct 
tree-level coupling and the other induced by loop effects.

The simplest example of the communication of supersymmetry breaking is
obtained in a theory in which the wall at $x^5 = \ell$ contains no boundary
matter fields.  We choose the gauge group to be $U(1)$ and write a 
a Fayet-Iliopoulos $D$ term on this boundary.  With the identification
of the $D$ term given in Section 2, the following boundary action 
preserves $N=1$ supersymmetry:
\begin{equation}
   {\cal L}_4  =  \kappa (X^3-\partial_5\Phi) \ .
\label{FIterm}\end{equation}
Integrating out the auxiliary field $X^3$ leads to a $\delta(0)$ term which
is an irrelevant constant.  The dynamical $\Phi$ field is affected by 
this term, in a manner that we can compute from the action
\begin{equation}
  S = \int d^5x \left\{ {1\over 2g^2} (\partial_M\Phi)^2 
              -  \kappa \partial_5 \Phi \delta(x^5-\ell) \right\} \ .
\label{Phiaction}\end{equation}
Varying this action with respect to $\Phi$, we find that the
 Fayet-Iliopoulos term induces a background expectation value of 
$\Phi$ which depends only
on $x^5$ and satisfies the equation
\begin{equation}
  {1\over g^2}\partial_5^2 \left\langle{\Phi} \right\rangle 
+  \kappa \partial_5 \delta(x^5 - \ell) = 0 \ .
\label{Phieq}\end{equation}
In solving this equation, we should remember that the geometry with 
mirror planes arose from a identification of points in a compactification
of $x^5$ on a circle.  Thus,  $\left\langle{\Phi(x^5)}\right\rangle$ 
must be  a periodic 
function of $x^5$ with period $2\ell$ and so $\partial_5\Phi$ must integrate
to zero around the circle. This boundary condition requires that we choose
as the solution to (\ref{Phieq})
\begin{equation}
    \partial_5\left\langle{\Phi}\right\rangle =
 - g^2 \kappa \left( \delta(x^5-\ell) - {1\over 2\ell}\right) \ .
\label{Phisol}\end{equation}
Inserting this result into the $D$-term coupling on the wall at $x^5 = 0$, 
given by (\ref{Xphi}), we find a scalar mass term
given by
\begin{equation}
      M_\phi^2 =   g^2 Q {\kappa\over 2\ell} \ ,
\label{Massval}\end{equation}
where $Q$ is the $U(1)$ charge of the scalar field,
with no corresponding mass term induced for the fermions on the wall.
If the dynamics on the wall at $x^5 = \ell$ gives a $D$-term of fixed
magnitude there, the $\Phi$ field transfers this across the
fifth dimension to create a soft scalar mass term on the wall at 
$x^5 = 0$.

One subtlety of the Fayet-Iliopoulos mechanism of supersymmetry breaking 
is that the symmetry breaking $D$ term can sometimes be compensated by
shifting the vacuum expectations value of a scalar field.  We can see
a similar possibility here.  Generalize the previous model to include
several chiral multiplets $\phi_i$ on the wall at $x^5 = 0$, and 
additional  chiral multiplets $\phi_j$ on the wall at $x^5 = \ell$.
(As always, it is important to distinguish between the boundary
scalar fields $\phi$ and the bulk field $\Phi$.)
Assign these multiplets the charges $Q_i$, $Q_j$ under the $U(1)$ symmetry.
In the most general situation, all of the scalar fields might acquire
vacuum expectation values.   Then the Lagrangian for $\Phi$ will take
the form
\begin{eqnarray}
  S &=& \int d^5x \bigg\{ {1\over 2g^2} 
          \left( (X^3)^2 + (\partial_M\Phi)^2 \right)
	+ (\sum_i Q_i \phi_i^\dagger \phi_i )
	(X^3-\partial_5\Phi)\delta(x^5) \nonumber \\ 
    & &  
      + (\kappa + \sum_j Q_j \phi_j^\dagger \phi_j)
(X^3 - \partial_5 \Phi )\delta(x^5-\ell) \bigg\} \ .
\label{morePhiaction}\end{eqnarray}
For simplicity, we  assume that the $\phi_i$ and $\phi_j$ are represented
only by vacuum expectation values that are independent of position on the
walls.  Then  varying the action (\ref{morePhiaction}) with respect to 
$\Phi$ gives an equation analogous to (\ref{Phieq}) whose solution is
\begin{equation}
    \partial_5\left\langle{\Phi}\right\rangle =  
	- g^2 \left[  (\sum_i Q_i \phi_i^\dagger \phi_i )
       \left( \delta(x^5) - {1\over 2\ell}\right) + 
( \kappa +   \sum_j Q_j \phi_j^\dagger \phi_j )
     \left( \delta(x^5-\ell) - {1\over 2\ell}\right) \right]\ .
\label{morePhisol}\end{equation}
This result reduces to (\ref{Phisol}) when we turn off the expection
values of the $\phi_i$ and $\phi_j$.
If we insert this expression into (\ref{morePhiaction}), and also integrate
out the auxiliary field $X^3$, the various $\delta(0)$ terms cancel, leaving
behind
\begin{equation}
  S = \int d^5x  \bigg\{
  - {g^2\over 4\ell} (\kappa + \sum_i Q_i \phi_i^\dagger \phi_i+ 
\sum_j Q_j \phi_j^\dagger \phi_j)^2
 \bigg\} \ .
\label{Phielim}\end{equation}
To minimize the vacuum energy, we must set the various vacuum expectation
values to the supersymmetric condition  
\begin{equation} 
 \kappa + \sum_i Q_i \phi_i^\dagger \phi_i+ 
\sum_j Q_j \phi_j^\dagger \phi_j = 0 \ ,
\label{susyc}\end{equation}
if this is possible.

If the supersymmetric theory on the wall at $x^5 = \ell $ breaks 
supersymmetry spontaneously without inducing a $D$ term, it is necessary to
go to a higher order in perturbation theory to find the supersymmetry-breaking
communication.  If supersymmetry breaking causes a mass splitting among 
chiral supermultiplets on the wall at $x^5 = \ell$, and these multiplets
couple to the gauge field in the bulk, then the diagrams shown in 
Figure~\ref{fig:twoloop} induce a supersymmetry-breaking mass for scalars
on the wall at $x^5 = 0$.  Since, in the scheme we are studying, the 
particle number of a chiral multiplet at $x^5 = 0$ is conserved, this is the
only soft supersymmetry-breaking term that can be generated.

The generation  of the scalar mass term  in this example is very similar to 
that in `gauge-mediated' 4-dimensional models of supersymmetry breaking
\cite{DNNS}.
The same set of diagrams appears, with only the difference that our 
gauge fields live in 5 dimensions. 

To illustrate the computation of these diagrams, we study the simplest 
multiplet which appears in models of gauge-mediation.  We introduce 
 on the wall at $x^5 = \ell$ two chiral superfields 
$(\phi', {\overline{ \phi}}')$
which transform
under the gauge group  as a 
vectorlike representation $(R' + \overline{ R}')$.  (Recall that we are using 
$R$ to denote 
representation of the chiral fields $\phi$ at $\ell = 0$.)
 We give this multiplet a 
supersymmetric mass $m$ and induce an additional mass term for the scalar
fields from the vacumm expectation value of an $F$-term.  Then the fermions
have a Dirac mass $m$, while the bosons have a (mass)$^2$ matrix
\begin{equation} 
      M^2 \pmatrix{\phi' \cr {\overline{\phi}}^{*\prime}\cr} = 
 \pmatrix{ m^2 & m^2x \cr m^2x & m^2}
 \pmatrix{\phi' \cr {\overline{\phi}}^{*\prime}\cr} \ .
\label{phimsqx}\end{equation}
The eigenvectors of this matrix
 are species $\phi_+^\prime$, $\phi_-^\prime$ in the 
representation $R'$.  Thus, we find the mass spectrum on the wall at 
$x^5 = \ell$, 
 \begin{equation}
 m^2(\phi^\prime_+) = m_+^2 \ , \qquad  m^2(\phi^\prime_-) = m_-^2 \ , \qquad
 m^2(\psi') = m^2 \ , 
\label{phimsq}\end{equation}
 with $m_\pm^2 = m^2 (1\pm x)$.  This spectrum satisfies
str$[M^2]= 0$.  Our calculation will follow closely the discussion of 
gauge-mediated scalar masses in this model given by Martin \cite{Martin}.
  It is straightforward to generalize our calculation to 
models of supersymmetry breaking with nonvanishing supertrace. 
 However,
in that case, the scalar masses induced by gauge-mediation are 
cutoff-dependent even in 4 dimensions \cite{PTr}. Similar divergences
appear also in the 5-dimensional case.

To compute the scalar mass, we repeat the calculation of the digrams
in Figure \ref{fig:twoloop}, now assigning to the particles in the
loop the mass spectrum described in the previous paragraph.
As in the previous section, the identity (\ref{repdeltazero}) can be
used to replace $(k^5)^2$ by $k^2$ in the numerator of the diagrams
with $\Phi$ exchange.  Then the result reduces to a sum of two-loop
scalar integrals, just as in the 4-dimensional case.

\begin{figure}
\begin{center}
\leavevmode
{\epsfxsize=2.00truein \epsfbox{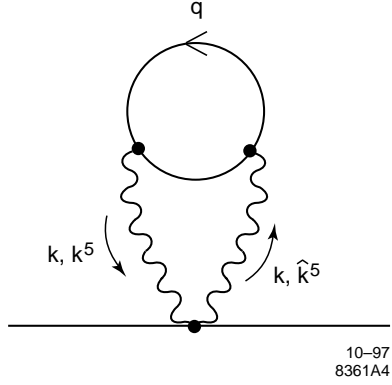}}
\end{center}
 \caption{The basic integral which appears in the two-loop contribution
    to the scalar field mass.}
\label{fig:basic}
\end{figure}

To write the result precisely, define  \cite{vdBV}
\begin{equation}
   (m_1 m_2| m_3 | m_4)  = \int {d^dk\over (2\pi)^d} \int {d^dq\over (2\pi)^d}
       {1\over k^2 + m_1^2}{1\over k^2 + m_2^2}
 {1\over q^2 + m_3^2}{1\over (q-k)^2 + m_4^2} 
\label{bracketdef}\end{equation}
to be the Euclidean (Wick-rotated)
scalar two-loop integral with four  propagators, and denote Euclidean
scalar integrals with additional or fewer propagators by brackets with 
more or fewer labels $m_i$.  In our calculation, 
$k^5$ is summed over the values $\pi n/\ell$; denote the sum in
 (\ref{inttwodef}) as
\begin{equation}
    \bigl[ {\cal A} \bigr] =  
 {1\over 2\ell}\sum_{n} \,
 {1\over 2\ell}\sum_{{\hat n}}  \  4 \cdot (-1)^{n + \hat n}
 {\cal A}(k^5,{\hat k}^5)  \ .
\label{fivebrac}\end{equation}
The basic scalar integral shown in Figure \ref{fig:basic} is then
written
\begin{equation}
           [ (k^5 {\hat k}^5 | m_1 | m_2)] \ . 
\label{basicint}\end{equation}
Finally, though a term with $k^2$ in the numerator can be reduced to 
scalar integrals, it is more convenient to retain this factor under the 
integral in (\ref{bracketdef}).  By abuse of notation, we will write a term
with $k^2$ in the numerator as, for example, 
 $[k^2 (k^5 {\hat k}^5 |m_2 | m_3)]$.

%%%%%%%%%%%%%%%%%%%%%%%%%%%%%%%%%%%%%%%%%%%%%%%%%%%%%%%%%%%%%%%%%%%%%%
%%%%%%%%%%%%%%%%%%%%%%%%%%%%%%%%%%%%%%%%%%%%%%%%%%%%%%%%%%%%%%%%%%%%%%

Using the notation, the scalar mass due to the diagrams of Figure
\ref{fig:twoloop} is 
\begin{equation}
m_\phi^2 = - g^4 C_2(R) C(R')  [ k^2 {\cal R} + 4 {\cal S} ] \ ,
\label{phiexpress}\end{equation}
where
\begin{eqnarray}
   {\cal R} &=& (k^5 {\hat k}^5 | m_+ | m_+)+  (k^5 {\hat k}^5 | m_- | m_-)
           +2  (k^5 {\hat k}^5 | m_+ | m_-)+4  (k^5 {\hat k}^5 | m | m)\nonumber \\ 
    & & -4  (k^5 {\hat k}^5 | m_+ | m) - 4
  (k^5 {\hat k}^5 | m_- | m)\nonumber \\ 
   {\cal S} &=& m_+^2\left\{(k^5 {\hat k}^5 | m_+ | m_+)-
              (k^5 {\hat k}^5 | m_+ | m) \right\}
 -  m^2\left\{(k^5 {\hat k}^5 | m | m)-
              (k^5 {\hat k}^5 | m_+ | m) \right\} \nonumber \\ 
          & &  + m_-^2\left\{(k^5 {\hat k}^5 | m_- | m_-)-
              (k^5 {\hat k}^5 | m_- | m) \right\}
 -  m^2\left\{(k^5 {\hat k}^5 | m | m)-
              (k^5 {\hat k}^5 | m_- | m) \right\} \ .
\label{mphianswer}\end{eqnarray}
This expression is full of cancellations which reflect the fact that the
answer vanishes when the mass spectrum is supersymmetric.
To evaluate this answer more explicitly,
 we must perform the sums over $k^5$ and $ {\hat k}^5$
and then carry out the two four-dimensional integrals.

The sums can be performed conveniently using a standard trick from 
finite temperature field theory.  Write a contour integral representation
\begin{equation}
  {1\over 2\ell}\sum_{n} \,2 (-1)^n\, {1\over k^2 + (k^5)^2}    
    = \oint {dk^5\over 2\pi} {2 e^{ik^5\ell}\over e^{2ik^5\ell}-1} 
 {1\over k^2 + (k^5)^2}  \ ,
\label{firstcontour}\end{equation}
where the contour encloses the poles of the integrand at $k^5 = \pi n/\ell$.
Draw the contour as a line from left to right just below the real axis and
another line from right to left just above this axis.  Push the first line
down and pick up the pole at $k^5 = -ik$; push the second line up and 
pick up the pole at $k^5 = ik$.  We find two identical contributions which
sum to 
\begin{equation}
            {1\over k}{1\over \sinh k\ell} \ . 
\label{thesinh}\end{equation} 
This manipulation can be performed separately on each of the propagators
joining the two walls.

At the same time, the scalar integrals over the momentum $q$ can be evaluated
explicitly.  Define the function $b(k^2,m_1^2,m_2^2)$ by 
\begin{equation}
    \int {d^d q\over (2\pi)^d}{1\over q^2 + m_1^2}{1\over (q-k)^2 + m_2^2}
 = {1\over (4\pi)^d}\big\{ {2\over \epsilon} - \gamma - b(k^2,m_1^2,m_2^2) + 
        {\cal O}(\epsilon)\big\}
\label{expandsc}\end{equation}
for $d = 4-\epsilon$.  When we evaluate the loop integrals on the 
wall in (\ref{mphianswer}), the divergent terms cancel and we are left
with differences  of these scalar  functions,
\begin{eqnarray}
   {\cal R} &\to&  R(k^2)= 
 b(k^2,m_+^2, m_+^2) + b(k^2,m_-^2, m_-^2)
 + 2 b(k^2,m_+^2, m_-^2) + 4 b(k^2,m^2, m^2) \nonumber \\ 
     & & -4  b(k^2,m_+^2, m^2) - 4 b(k^2,m_-^2, m^2) \nonumber \\ 
    {\cal S} &\to&  S(k^2)= 
 m_+^2 \{ b(k^2,m_+^2, m_+^2) - b(k^2,m_+^2, m^2)\}
          - m^2 \{ b(k^2,m^2, m^2) - b(k^2,m_+^2, m^2)\}\nonumber \\ 
     & &   + m_-^2 \{ b(k^2,m_-^2, m_-^2) - b(k^2,m_-^2, m^2)\}
          - m^2 \{ b(k^2,m^2, m^2) - b(k^2,m_-^2, m^2)\} \ .
\label{thebees}\end{eqnarray}
If we then define 
\begin{equation}
    P(k^2) = k^2 R(k^2) + 4 S(k^2) \ ,
\label{thehh}\end{equation}
then the combination of these two tricks brings (\ref{phiexpress}) into the
form of an integral over $k$.  Since this integral is spherically symmetric,
we can replace  $d^4k = 2\pi^2 dk k^3$ and write  (\ref{phiexpress}) as
\begin{equation} 
  m_\phi^2 = 2 \left({g^2\over (4\pi)^2}\right)^2 C_2(R) C(R') \int^\infty_0
       dk k {1\over \sinh^2 k\ell} P(k^2)  \ . 
\label{thePform}\end{equation}

The function $P(k^2)$ is elementary, and it is not difficult to work out 
its asymptotic behavior for large and for small $k^2$. 
 We present these formulae in the Appendix.
It is relevant that $P(k^2) \sim k^2$ as $k^2 \to 0$, so that $P(k^2)$ is a
field-strength renormalization \cite{GRat}
(as the notation is meant to suggest).
 As $k^2 \to \infty$, $P(k^2) \sim 
\log(k^2/m^2)/k^2$.  With this information, one can work out the asymptotic
behaviors of $m_\phi^2$.

For small $\ell$, we might expect to go back the the 4-dimensional situation.
Formally, taking $\ell\to 0$ in (\ref{thesinh}) turns this expression into
\begin{equation}
            {2\over 2\ell} {1\over k^2} \ ,
\label{theellzero}\end{equation}
which is the $k^5=0$ term in the sum (\ref{firstcontour}). Using the explicit
asymptotic behavior of $P(k^2)$, we can see that the integral (\ref{thePform})
remains well-defined in this limit. Thus, the manipulation is permitted
and the result for $m_\phi^2$ becomes just the 4-dimensional gauge-mediation
result with the replacement
\begin{equation}
      \left({g^2\over (4\pi)^2}\right)^2 \to  
       \left({g^2\over (4\pi)^2}\right)^2  {1\over \ell^2} \ .
\label{fourreplace}\end{equation}
We will write out this result explicitly below. Note that $g^2/\ell$ is the 
effective 4-dimensional gauge coupling obtained by simple dimensional 
redution.  

Another way to derive this result is to show that, for $\ell \to 0$,
all terms in the sum over $k^5$ and ${\hat k^5}$ are explicitly 
suppressed by the factor $\ell^2$ except for the term with 
 $k^5 = {\hat k^5}= 0$.  Again, the asymptotic behavior $P(k^2) \sim 1/k^2$
is necessary to complete this argument.

For large $\ell$. the hyperbolic sine in the denomination of (\ref{thePform})
cuts off the integrand at very small $k$.  Thus, we can find the asymptotic
behavior by replacing $P(k^2)$ by its leading term for small $k^2$, which
is proportional to $k^2$.  This gives a result proportional to 
\begin{equation}
  \left({g^2\over (4\pi)^2}\right)^2  {1\over \ell^4}\ .
\label{thelargel}\end{equation}

%%%%%%%%%%%%%%%%%%%%%%%%%%%%%%%%%%%%%%%%%%%%%%%%%%%%%%%%%%%%%%%%%%%%%%
\begin{figure}
\begin{center}
\leavevmode
{\epsfxsize=4.00truein \epsfbox{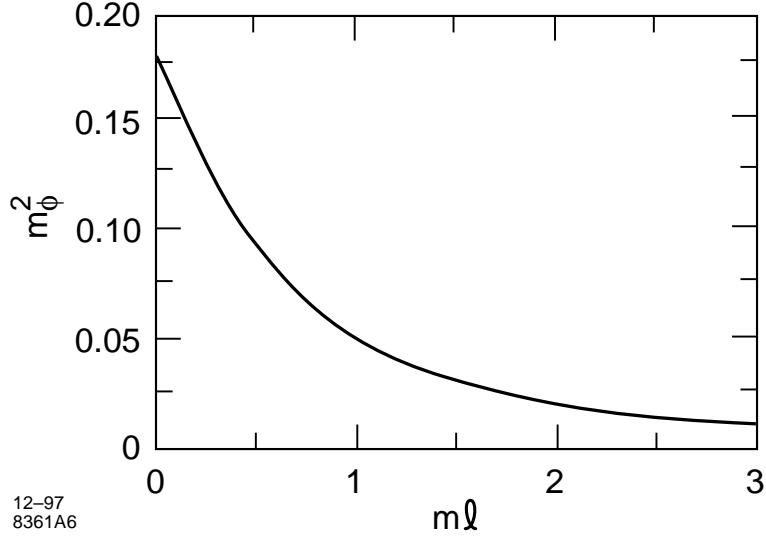}}
\end{center}
 \caption{Behavior of the induced supersymmetry breaking mass for 
scalars at $x^5 = 0$ as a function of $\ell$.  We plot $m_\phi^2$
in units of $2 C_2(R) C(R')
(g^2/(4\pi)^2)^2 )\cdot (m^2/\ell^2))$.}
\label{fig:themass}
\end{figure}
%%%%%%%%%%%%%%%%%%%%%%%%%%%%%%%%%%%%%%%%%%%%%%%%%%%%%%%%%%%%%%%%%%%%%%

Working out all of the details (with the help of some formulae from the 
Appendix), we find, as $m \ell \to 0$, the 4-dimensional
 form~\cite{Martin,DGP} 
\begin{eqnarray}
  m_\phi^2 &=& 2C_2(R)C(R') \left({g^2\over (4\pi)^2}\right)^2 \cdot 
 {m^2\over \ell^2} \nonumber \\ 
    & & \cdot\left\{ 2(1+x)\left[\log(1+x) - 2 \mbox{\rm Li}_2({x\over 1+x})
                + \frac{1}{2} \mbox{\rm Li}_2({2x\over 1+x})\right]
 + (x \leftrightarrow -x)\right\}\ ,
\label{thesmalllmp}\end{eqnarray}
where $\mbox{\rm Li}_2(x)$ is the dilogarithm, and,  as $m \ell \to \infty$,
\begin{eqnarray}
  m_\phi^2 &=& 2C_2(R)C(R') \left({g^2\over (4\pi)^2}\right)^2 \cdot
 {1\over \ell^4} \cdot \zeta(3)\nonumber \\ 
  & &  \cdot\left\{{3\over 2}\left[{4 + x - 2x^2\over x^2}\log(1+x)
      -  {4-x\over x}\right]  + (x \leftrightarrow -x)\right\}\ .
\label{thelargemp}\end{eqnarray}
In both of these expressions, the quantity in brackets tends to $x^2$ as
$x\to 0$. We see that the induced soft supersymmetry breaking mass term
crosses over from the 4-dimensional behavior to a smaller functional form
as $\ell$ becomes larger than $1/m$.
In Figure \ref{fig:themass}, we graph the form of the mass term as a function
of $\ell$ for the illustrative value $x = 0.3$.

There is another way to understand the behavior of the scalar mass term
for $m\ell$ large.  If $m$ is large, we can consider the inner loop in 
Figure \ref{fig:basic} to contract to a point. More precisely, because the
function $P(k^2)$ is proportional to $k^2$ when $k$ is small, this loop
gives the pointlike operator $(-\nabla^2)$ acting on the two propagators
which run from one wall to the other.  The remaining one-loop integral
may be evaluated in Euclidean coordinate space.  There is one small 
subtlety to note.  The representation of (\ref{fiveprop}) in Euclidean
coordinate space is 
\begin{eqnarray}
  \left\langle{ a(x,x^5) a(y,y^5)}\right\rangle &=
	& {1\over 8\pi^2}\sum_m \biggl(
 {1\over [ ( x-y)^2 + (x^5 - y^5 + 2m\ell)^2 ]^{3/2} }\nonumber \\ 
 & &   + 
 P {1\over [ ( x-y)^2 + (x^5 + y^5 + 2m\ell)^2 ]^{3/2} } \biggr)\ ,
\label{fivepropC}\end{eqnarray}
with $m$ summed over all integers.
When this expression is evaluated with one end at $x^5 = \ell$ and the other
at $y^5 = 0$, we find (for $P=+1$) 
\begin{equation}
  \left\langle{ a(x,\ell) a(0,0)}\right\rangle =   {1\over 8\pi^2}\sum_m 
 {2\over [ x^2 + (2m+1)^2\ell^2 ]^{3/2} } \ .
\label{fivepropCh}\end{equation}
Then the evaluation of $m_\phi^2$ involves the expression
\begin{equation}
   I= \sum_{m,\hat m} \int d^4x {2\over 8\pi^2 [x^2 + (2m+1)^2\ell^2]^{3/2}}
 (-\nabla^2)  {2\over 8\pi^2 [x^2 + (2\hat m+1)^2\ell^2]^{3/2}} \ ,
\label{thecoordexp}\end{equation}
containing two propagators which run from
a point 0 on the wall at $x^5 = 0$ to a point $x$ on the wall at $x^5 = 
\ell$.   By combining the
two denominators with a Feynman parameter, it is not difficult to do the
integral explicitly and then sum over $m$ and $\hat m$.  The result is
\begin{equation}
  I = {3\over 16\pi^2} \zeta(3) {1\over \ell^4} \ .
\label{theIresult}\end{equation}
Multiplying this by the coefficient of $k^2$ in $P(k^2)$, we find again
the result (\ref{thelargemp}).   This presentation explains the physical 
origin of the $1/\ell^4$ behavior of the diagrams.

\section{Casimir energy}

At the same time that supersymmetry breaking on the wall at $x^5 = \ell$
induces soft super\-sym\-me\-try-breaking terms in other parts of the theory,
it also creates  a nonzero vacuum energy.  We are particularly interested
in the part of this energy which depends on $\ell$---the Casimir
energy~\cite{Casimir}---since this term will eventually form a part of 
the balance  which determines the physical value of $\ell$.  In this section,
we will compute the Casimir energy due to the two mechanisms of supersymmetry
breaking discussed in the previous section.  We find it interesting that 
these calculations run almost in parallel to the calculations of the 
induced scalar mass term.

Consider first the case of a Fayet-Iliopoulos $D$-term on the boundary
at $x^5 = \ell$.  The coupling of this term to the bulk fields is 
described by the Lagrangian (\ref{Phiaction}), plus a term proportional
to $\delta(0)$ resulting from integrating out $X^3$.  Since
(\ref{Phiaction}) 
is quadratic in $\Phi$, we can integrate this field out explicitly.  Using
the propagator (\ref{Phisol}), the coupling of $\Phi$ to the boundary 
leads to 
\begin{equation}
  S_{\scriptscriptstyle eff} = 
	\int d^5 x \delta(x^5-\ell) \cdot \frac{1}{2} \kappa
 \left(-{g^2 \kappa\over 2\ell}\right) \ ,
\label{thenewS}\end{equation}
plus an $\ell$-independent term proportional to $\delta(0)$.  Thus, we 
find for the Casimir energy per 4-dimensional volume,
\begin{equation}
   E_C/V_4 =  + {g^2 \kappa^2\over 4\ell} \ .
\label{theCasone}\end{equation}

If there are $D$-terms on both boundaries, or if the fields $\phi_i$ on
the two boundaries obtain expectation values as in (\ref{morePhiaction}), 
the sum  of the two $D$ terms appears in place of $\kappa$ in 
(\ref{theCasone}).   If the two $D$ terms are equal and opposite,
 the Casimir energy
vanishes.  Also, as we observed already in (\ref{Phielim}),
the $\delta(0)$ terms from 
integrating out $X^3$ and $\Phi$ precisely cancel.  Thus, in this case,
the vacuum energy remains just at zero, as expected from the supersymmetry
of the situation.

%%%%%%%%%%%%%%%%%%%%%%%%%%%%%%%%%%%%%%%%%%%%%%%%%%%%%%%%%%%%%%%%%%%%%%
\begin{figure}
\begin{center}
\leavevmode
{\epsfxsize=4.00truein \epsfbox{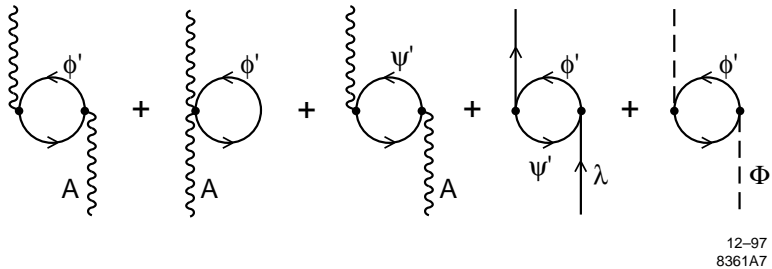}}
\end{center}
 \caption{Feynman diagrams contributing to the Casimir energy
due to loop effects of the supermultiplet on the wall at $x^5 = \ell$.}
\label{fig:Cloops}
\end{figure}
%%%%%%%%%%%%%%%%%%%%%%%%%%%%%%%%%%%%%%%%%%%%%%%%%%%%%%%%%%%%%%%%%%%%%%

In the case in which the spectrum at $x^5 = \ell$ violates supersymmetry
but there is no induced $D$ term, the Casimir energy must be generated by
radiative corrections.  The leading contribution comes from the diagrams
shown in Figure \ref{fig:Cloops}.  These diagrams involve a closed loop 
on the boundary at $x^5 = \ell$ and a propagator from the 5-dimensional
Yang-Mills theory which winds around the compactified direction.

Though perhaps it is not completely obvious from the beginning, the 
structure of the diagrams in Figure \ref{fig:Cloops} is very similar to 
that of the diagrams in Figure \ref{fig:twoloop}.  As in the previous 
section, we will describe the calculation for the case str$[M^2]= 0$.

In the diagrams of Figure \ref{fig:Cloops}, we have only one sum over $k^5$. 
Thus,   define for this section
\begin{equation}
    \bigl[ {\cal B} \bigr] =  
 {1\over 2\ell}\sum_{n} \, {\cal B}(k^5) \ .
\label{fivebractwo}\end{equation}
Then the Casimir energy resulting from this set of diagrams can be written
in terms of Euclidean scalar two-loop integrals as
\begin{equation}
E_C/V_4 = \frac{1}{2} g^2 d_G C(R')  [ k^2 {\cal R_C} + 4 {\cal S_C} ] \ ,
\label{Eexpress}\end{equation}
where  $d_G C(R') = {\mbox{\rm tr}}_{R'} [t^A t^A]$, and 
\begin{eqnarray}
   {\cal R_C} &=& (k^5| m_+ | m_+)+  (k^5| m_- | m_-)
           +2  (k^5 | m_+ | m_-)+4  (k^5| m | m)\nonumber \\ 
    & &  -4  (k^5 | m_+ | m) - 4
  (k^5 | m_- | m)\nonumber \\ 
   {\cal S_C} &=& m_+^2\left\{(k^5 | m_+ | m_+)-
              (k^5 | m_+ | m) \right\}
 -  m^2\left\{(k^5 | m | m)-
              (k^5 | m_+ | m) \right\} \nonumber \\ 
          & & m_-^2\left\{(k^5 | m_- | m_-)-
              (k^5 | m_- | m) \right\}
 -  m^2\left\{(k^5 | m | m)-
              (k^5 | m_- | m) \right\} \ .
\label{Eanswer}\end{eqnarray}
The inner loop of each of these two-loop integrals can be evaluated 
explicitly, giving the same functions $R(k^2)$, $S(k^2)$, $P(k^2)$
that we saw earlier in (\ref{thebees}) and (\ref{thehh}).

Again we can simplify the sum over $k^5$ using the contour trick from 
finite temperature field theory.  Write
\begin{equation}
  {1\over 2\ell}\sum_{n} \, {1\over k^2 + (k^5)^2}    
    = \oint {dk^5\over 2\pi}{1\over e^{2ik^5\ell}-1} 
 {1\over k^2 + (k^5)^2}  \ ,
\label{secondcontour}\end{equation}
where the contour encloses the poles of the integrand at $k^5 = \pi n/\ell$.
Draw the contour as a line from left to right just below the real axis and
another line from right to left just above this axis. Push the first line
down  and pick up the pole at  $k^5 = -ik$.
For the 
contour integral on the line above the axis, replace
\begin{equation}
   {1\over e^{2ik^5\ell}-1}  =  - 1 - {1\over e^{-2ik^5 \ell} -1} \ .
\label{theswitch}\end{equation}
In the second term, push the contour up and pick up the pole at
 $k^5 = ik$. These manipulations convert (\ref{secondcontour}) to the form
\begin{equation}
       {1\over k} {1\over e^{2k\ell} -1}
  + \int^\infty_{-\infty} {dk^5\over (2\pi)} \ .
\label{theswitchdone}\end{equation}
The second term in (\ref{theswitchdone}) is independent of $\ell$; it 
represents the contribution to the vacuum energy 
 of the 4-dimensional wall  in an infinite 5-dimensional volume.
Equivalently, from the point of view of propagators in coordinate space,
this term is the contribution of the propagators that go from the 
wall back to the wall without winding around $x^5$.  In any event, this 
term does not contribute to the Casimir energy, and we may drop it.

After these manipulations, the Casimir energy (\ref{Eexpress}) takes the form
\begin{equation} 
 E_C/V_4 = 
	- \frac{1}{2} \left({g^2\over (4\pi)^4}\right) d_G C(R') \int^\infty_0
       dk k^2  {1\over e^{2k\ell} -1} P(k^2)  \ , 
\label{theEPform}\end{equation}
where $P(k^2)$ is the same field strength renormalization function that
appeared in (\ref{thePform}).

As in the previous section, we can analyze the two-loop integral in the 
limits of small and large $\ell$.  Consider first the limit $\ell\to 0$. 
If we formally take the limit of small $\ell$ in (\ref{theEPform}), we 
obtain a divergent integral
\begin{equation} 
    E_C/V_4 \sim  -  \int^\infty_0 dk {k\over \ell} {1\over k^2}\log k^2 \ .
\label{firstestE}\end{equation}
Thus, unlike the case of $m_\phi^2$, the contributions to the Casimir 
energy are dominated by large values of $k^2$.  The estimate
\begin{equation}
      \int^\infty_0 dk k^2 {1\over e^{2k\ell} -1}  {1\over k^2}\log k^2 
   \sim  {1\over 2\ell} \log^2{1\over m\ell} 
\label{estimateE}\end{equation}
and the asymptotic formula for $P(k^2)$ given in the Appendix gives a 
precise formula for the small $\ell$ behavior:
\begin{equation}
E_C/V_4 =  
	- \frac{1}{2} \left({g^2\over (4\pi)^4}\right) d_G C(R') 
	\cdot {4 m^4 x^2 
       \over \ell}
        \log^2 {1\over m\ell} \ .
\label{smalllE}\end{equation}

%%%%%%%%%%%%%%%%%%%%%%%%%%%%%%%%%%%%%%%%%%%%%%%%%%%%%%%%%%%%%%%%%%%%%%
\begin{figure}
\begin{center}
\leavevmode
{\epsfxsize=4.00truein \epsfbox{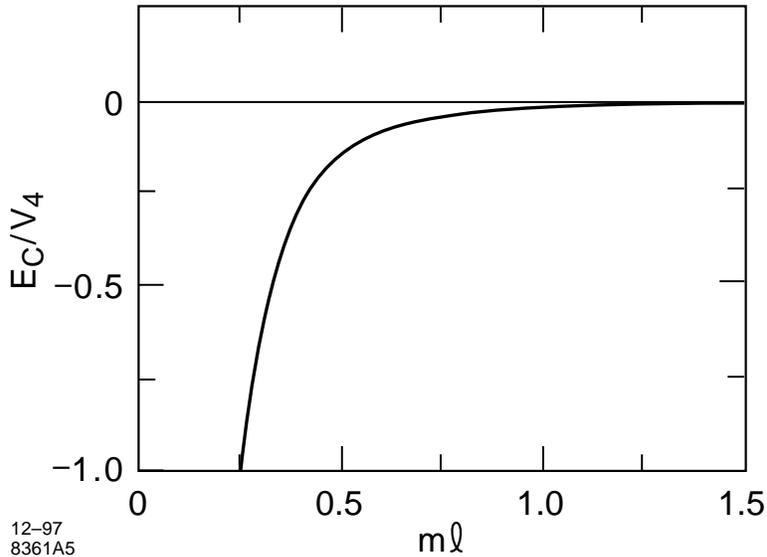}}
\end{center}
 \caption{Behavior of the Casimir energy as
 a function of $\ell$.  We plot $(E_C/V_4)$ in units of ${1\over 2}  d_G C(R')
(g^2/(4\pi)^4))$.}
\label{fig:theenergy}
\end{figure}
%%%%%%%%%%%%%%%%%%%%%%%%%%%%%%%%%%%%%%%%%%%%%%%%%%%%%%%%%%%%%%%%%%%%%%

For large $\ell$, the analysis can be done along the same lines as those
we used for $m_\phi^2$.  The denominator of (\ref{theEPform}) cuts off
the integrand for small $k^2$.  Thus, we can replace $P(k)$ by  its
leading term as $k^2 \to 0$.  With this approximation, the integral is 
easily evaluated, and we obtain
\begin{eqnarray}
E_C/V_4 &=&  - \frac{1}{2}
 \left({g^2\over (4\pi)^4}\right) d_G C(R') \zeta(5) {1\over 
              \ell^5} \nonumber \\ 
     & &  \left\{{3\over 2}\left[{4 + x - 2x^2\over x^2}\log(1+x)
      - 
           {4-x\over x}\right]  + (x \leftrightarrow -x)\right\}\ .
\label{largelE}\end{eqnarray}
Comparing (\ref{smalllE}) and (\ref{largelE}), we see the same crossover
that we found previously from (\ref{thesmalllmp}) to (\ref{thelargemp}).
In Figure \ref{fig:theenergy}, we graph the form of the Casimir energy
 as a function
of $\ell$ for the illustrative value $x = 0.3$.

As in the previous section, the behavior of the Casimir energy for large
$\ell$ is studied most easily in Euclidean coordinate space.  If $\ell$
or $m$ is large, the inner loop of each two-loop diagram can be 
contracted to a local operator proportional to  $(-\nabla^2)$.
Then the Casimir energy is proportional to an expression in which this
operator acts on a propagator which runs around the compact 
direction.  More specifically, the Casimir energy is proportional to 
\begin{equation}
J =  \sum_{m} 
 (-\nabla^2)  {1\over 8\pi^2 (x^2 + (m\ell)^2)^{3/2}} \biggr|_{x=0}\ ,
\label{thecoordEexp}\end{equation}
where the sum over $m$ runs over all integers except $m=0$.  This 
expression evaluates to 
\begin{equation}
J = {3\over 32\pi^2} \zeta(5) {1\over \ell^5} \ .
\label{theJval}\end{equation}
Multiplying this by the coefficient of $k^2$ in $P(k^2)$, we return to 
the result (\ref{largelE}). 

Both of the individual contributions to the Casimir energy that we found
in this section are monotonic in $\ell$.  We find it interesting, though, 
that these two contributions have opposite signs.  Thus, it is possible
that, in a realistic theory, we could find balancing contributions to the
Casimir energy that stabilize the value of $\ell$ at a nonzero
value.

\section{Ho\v{r}ava's supersymmetry-breaking structure}

Now that we have analyzed mechanisms for supersymmetry breaking in our
toy model, it is interesting to ask whether this sheds light on the 
mechanism of supersymmetry breaking in string theory proposed by 
Ho\v{r}ava \cite{allelse}.  We can see the connection by making a dimensional
reduction of Ho\v{r}ava's system from 11 to 5 dimensions, taking the
compact 6 dimensions to be a Calabi-Yau manifold.  Under this 
reduction, the boundary gaugino condensate becomes a 4-dimensional 
scalar gaugino bilinear on the boundary.  The relevant components of the 
3-form gauge field in the bulk are those that multiply the the (3,0) or
(0,3) forms of the Calabi-Yau space,
\begin{equation} 
        C_{ABC} (x,x^5,y) =   c(x,x^5) \Omega_{ABC}(y) + \cdots
\label{reduceC}\end{equation}
These components form two complex 5-dimensional fields which belong to a 
hypermultiplet in the bulk.  Thus, we can try to recover Ho\v{r}ava's 
coupling of the bulk and boundary fields by considering the coupling
of a hypermultiplet in the bulk to chiral fields on the boundary.

We can analyze this problem using arguments similar to those in Section 2.
The five-dimensional hypermultiplet 
consists of a pair of complex scalars $A^i$, a Dirac fermion $\chi$,
and a pair of complex auxiliary fields $F^i$.  Under supersymmetry
they transform as follows \cite{basicsusy}:
\begin{eqnarray}
{\delta_{\xi}} A^i & = &  -\sqrt{2}\epsilon^{ij}\overline{{\xi}}^j \chi \nonumber \\ 
{\delta_{\xi}} \chi   & = 
	&  +\sqrt{2} i \gamma^M \partial_M A^i \epsilon^{ij} \xi^j 
			+ \sqrt{2} F^i \xi^i   \nonumber \\ 
{\delta_{\xi}} F^i  & = 
	&  -\sqrt{2} i \overline{{\xi}}^i \gamma^M \partial_M \chi  \ .
\label{hypertrans}\end{eqnarray}

To carry out the orientifold projection, we must consistently assign 
parities $P$ to the various fields and impose the boundary conditions
(\ref{parityfive}).   Here is a consistent set of assignments which 
gives $N=1$ 
supersymmetry on the wall:
\begin{equation}
\begin{tabular}{|c|c|c|}                     
  & \ $P=+1$ \ & \ $P=-1$ \ \\ \hline\hline
$\xi$       & $\xi^1_L$     & $\xi^2_L$     \\ \hline
$A^i$       & $A^1$         & $A^2$        \\ \hline
$\chi$      & $\chi_L$      & $\chi_R$      \\ \hline
${\rm F}^i$ & ${\rm F}^1$   & ${\rm F}^2$   \\ \hline
\end{tabular}
\label{hypertable}\end{equation}
As in Section 2, we project out the 
 odd-parity states and consider the 
 supersymmetry on the boundary generated by  $\xi^1_L$.  The 
transformations (\ref{hypertrans}) specialize to 
\begin{eqnarray}
{\delta_{\xi}} A^1         & = &  \sqrt{2} \xi_L^{1T}   \chi_L \nonumber \\ 
{\delta_{\xi}} \chi_L      & = 
	&  \sqrt{2} i \sigma^m \partial_m A^1  \xi_L^{1*}  
		-  \sqrt{2} \partial_5 A^2 \xi_L^1  
		+ \sqrt{2} F^1 \xi_L^1  \nonumber \\ 
{\delta_{\xi}} F^1   & = &  i \sqrt{2} \xi_L^{1 \dagger} \overline{\sigma}^m
			\partial_m \chi_L + \sqrt{2} 
			\xi_L^{1 \dagger} \partial_5\chi_R
       \nonumber \\ 
{\delta_{\xi}} \partial_5 A^2  & = 
	& \sqrt{2} \xi_L^{1 \dagger} \partial_5 \chi_R \ .
\label{wallhyper}\end{eqnarray}
These transformations imply that 
\begin{equation} 
{\delta_{\xi}} (F^1 - \partial_5 A^2) = \sqrt{2} i \xi_L^{1 \dagger}
			\overline{\sigma}^m \partial_m  \chi_L \ ,
\label{Findent}\end{equation}
Then $A^1$, $\chi_L$, $(F^1 - \partial_5 A^2)$ transform as the 
 complex scalar, chiral fermion, and auxiliary field components of a
 four-dimensional $N=1$  chiral multiplet.

We can use this set of fields to write a coupling of the bulk 
hypermultiplet to chiral superfields on the boundary.  In particular, 
the boundary theory might have a superpotential which depends on the
boundary chiral fields $\phi_i$ and the boundary value of the field 
$A^1$.  The superpotential term then includes the boundary action
\begin{equation}
      {\cal L}_4 =   (F^1-\partial_5 A^2) {dW\over d A^1} + \cdots\ .
\label{LfourforAone}\end{equation}
If we integrate out the auxiliary field $F^1$ and write the resulting 
action in 5 dimensions, we arrive at the structure
\begin{equation} 
    {\cal L}_5 = 
	|\partial_M A^2|^2  -  \delta(x^5) 
	\left[\partial_5 A^2  {dW\over d A^1}
           + h.c. \right] + (\delta(x^5))^2 \left|   {dW\over d A^1}\right|
\label{LfiveforA}\end{equation}
If we indentify $A^2$ with the scalar component of $C_{ABC}$ shown in 
(\ref{reduceC}) and $(dW/dA^1)$ with the $E_8$ gaugino condensate, this 
reproduces the perfect-square structure (\ref{Hsform}) found by 
Ho\v{r}ava \cite{hw2,allelse}.
 
From here, we could go on to discuss the communication of supersymmetry
breaking. If we simply assume a fixed value of the gaugino condensate
and solve for $A^2$ as  in (\ref{Phisol}), we find a universal gaugino
mass proportional to $1/\ell$, as in \cite{LLN,NOY,LT}. This leads to 
conventional
supergravity-mediated supersymmetry breaking scenario.
 It would be 
very interesting to know whether there are other  possibilities.
In particular, it would be interesting to find
 a perturbative hierarchy of soft supersymmetry-breaking terms 
similar to the one that we discussed in Section 4.  To search for such 
possibilities,
it is necessary to understand the general coupling  of 
boundary matter fields to supergravity. We are currently investigating that
question.

\section{Conclusion}

In this paper, we have shown how easy it is to construct consistent 
couplings of five-dimensional supermultiplets to matter multiplets on 
orientifold walls by analyzing the transformation properties of the 
associated auxiliary fields.  We applied this method to some simple
models with bulk and boundary fields and exhibited several possibilities
for the communication of supersymmetry breaking from one wall to the other.
We hope that this method will generalize to supergravity and allow 
a more complete understanding of the supersymmetry breaking and its 
phenomenology in the Ho\v{r}ava-Witten approach to unification.

  \Acknowledgements

We are grateful to our colleagues at SLAC and Stanford for many discussions
of the issues of this paper, and to Michael Dine, Costas Kounnas,
Steve Martin,
Hans Peter Nilles, 
Erich Poppitz, Arvind Rajaraman, Lisa Randall, and Leonard Susskind for 
useful suggestions.  This work was supported by the Department of Energy 
under Contract No. DE--AC03--76SF00515.

\appendix

\section*{More about the two-loop self-energy}

In this appendix, we will give some further details of the two-loop
self-energy calculations discussed in Sections 3 and 4.

As we explained in (\ref{twoloopform}), our strategy for computing the 
diagrams shown in Figure~\ref{fig:twoloop} began with bringing each diagram 
into the form 
\begin{equation}
  M^2 = 
 - g^4 C_2(R)  C(R') \int_q \int_{k55}{ N(k,k^5,{\hat k}^5,q)\over 
 (k^2)(k^2 - (k^5)^2)(k^2 -( {\hat k}^5)^2) (q^2-m_1^2)((q-k)^2-m_2^2)} 
\label{twoloopformagain}\end{equation}
for the $m_1$, $m_2$ appropriate to the inner loop of the diagram.
We now give 
the contributions of the various diagrams to the numerator polynomial 
$N$.  In the following formula, we write the contributions to 
$N$ as a sum, following the order of the diagrams
 in  Figure~\ref{fig:twoloop}, although
properly each separate
term should receive the appropriate particle masses in the
denominator.  The expression is given for the mass spectrum (\ref{phimsq})
considered in Section 4; for the analysis of Section 3, one should 
set all masses
equal to zero.  With this explanation,
\begin{eqnarray}
N &=& 2(k\cdot(2q-k))^2 - 2(q^2- m_+^2 + q^2 -  m_-^2) 
	k^2 - 2 (2q-k)^2 k^2\nonumber \\ 
 & &  + 8 (q^2- m_+^2 + q^2 -  m_-^2) k^2 
 +4(q\cdot(q-k) k^2 - 2q\cdot k (q-k)\cdot k - m^2k^2)\nonumber \\  & &
 -8 k^2 (q\cdot (q-k)
      -2m^2)
       + 16 k^2 k\cdot (q-k) + 2k^2 + 0 + 0  \ .
\label{Nvalue}\end{eqnarray}
If we set all masses equal to zero, this expression vanishes after the use
of the $q\leftrightarrow (k-q)$ symmetry of the denominator. 
With nonzero masses, some simple rearrangements and a Euclidean
rotation bring the expression for $m_\phi^2$ into the form
(\ref{phiexpress}).

In our analysis of (\ref{phiexpress}), we made use of the self-energy integral
$b(k^2,m_1^2,m_2^2)$ defined by (\ref{expandsc}).  We can write $b$ more
explicitly as
\begin{eqnarray}
b(k^2,m_1^2,m_2^2) &=& \int^\infty_0 dx \log\left(x(1-x)k^2 + xm_1^2 + (1-x)
  m_2^2\right)\nonumber \\ 
 & = & A \log\left[{(A + B_1)(A+ B_2)\over (A-B_1)(A-B_2)}\right]
        + B_2 \log m_1^2 + B_1\log m_2^2 - 2  \ ,
\label{valofb}\end{eqnarray}
where 
\begin{equation}
 A = \left[{k^4 + 2k^2(m_1^2+m_2^2) + (m_1^2-m_2^2)^2\over 4k^4}\right]^{1/2}
\label{Avalue}\end{equation}
and
\begin{equation}
 B_1 = {k^2 + m_1^2 - m_2^2\over 2k^2}\ , \qquad 
 B_2 = {k^2 + m_2^2 - m_1^2\over 2k^2}\ . 
\label{Bvalue}\end{equation}
From $b(k^2, m_1^2, m_2^2)$, we can compute the combinations $R(k^2)$,
$S(k^2)$, $P(k^2)$ defined in (\ref{thebees}) and (\ref{thePform}).
We evaluate these expressions using the mass spectrum derived from 
(\ref{phimsq}).
It is straightforward to work out the asymptotic behavior of these 
functions for large and small values of $k^2$.  For $P(k^2)$, we find,
as $m^2 k^2 \to 0$, 
\begin{equation}
P(k^2) = k^2 \left[ {4 + x - 2x^2\over x^2 }\log (1+x) + 1 + (x\leftrightarrow
        -x) \right] + {\cal O}(k^4) \ .
\label{smallkP}\end{equation}
and as $m^2 k^2 \to \infty$,
\begin{equation} 
P(k^2) = {4m^4\over k^2}\left[x^2 \log{k^2\over m^2}
 - (x^2 + 3x +2)\log(1+x) -  x^2  + (x\leftrightarrow
        -x )\right] + {\cal O}(k^{-4}) \ .
\label{largekP}\end{equation}

The computation of the Casimir energy reported in Section 5 is very similar
to the computation of $m_\phi^2$ and, in particular, uses the same 
auxiliary function $P(k^2)$.


\begin{thebibliography}{99}



\bibitem{hw1} P. Ho\v{r}ava and E. Witten,  {{\em Nucl. Phys.}
B} {\bf 460}, 506 (1996),
 hep-th/9510209.

\bibitem{hw2} P. Ho\v{r}ava and E. Witten,  {{\em Nucl. Phys.}
B} {\bf 475}, 94 (1996),
 hep-th/9603142.

\bibitem{unification} E. Witten,  {{\em Nucl. Phys.} B} {\bf 471}, 135 (1996),
  hep-th/9602070. 

\bibitem{allelse} P. Ho\v{r}ava,  {{\em Phys. Rev.} D} {\bf 54}, 7561 (1996),
 hep-th/9608019.

\bibitem{sharpe}
E. Sharpe,  hep-th/9611196.

\bibitem{Casimir}
 H. G. B. Casimir, {\em Proc. Kong. Ned. Akad. Wet.} B {\bf 51}, 793 (1948).

\bibitem{Banks} T. Banks and M. Dine,  {{\em Nucl. Phys.}
B} {\bf 479}, 173 (1996),
 hep-th/9605136.

\bibitem{ABF}  
I. Antoniadis, S.  Ferrara, and T. R. Taylor,  {{\em Nucl. Phys.}
B} {\bf 460}, 489 (1996),
 hep-th/9511108.

\bibitem{Choi}  K. Choi,  {{\em Phys. Rev.} D} {\bf 56}, 6588 (1997),
 hep-th/9706171.

\bibitem{LLN}
T.-J. Li, J. L. Lopez, and D. V. Nanopoulos,
    {{\em Phys. Rev.} D} {\bf 56}, 2602 (1997),  hep-ph/9704247.

\bibitem{NOY}
H. P. Nilles, M. Olechowski, and M. Yamaguchi, {{\em Phys. Lett.} B} {\bf 415}, 24 (1997), hep-th/9707143.

\bibitem{LT}
Z. Lalak and S. Thomas,  {{\em Nucl. Phys.}
B} {\bf 515}, 55 (1998), hep-th/9707223.


\bibitem{ChoiandM}
K. Choi, H. B. Kim, and C. Munoz, hep-th/9711158.

\bibitem{Ovrut}
 A. Lukas, B. A. Ovrut, and D. Waldram, hep-th/9711197.

\bibitem{DSRW}
M. Dine, R. Rohm, N. Seiberg,  and E. Witten, 
    {{\em Phys. Lett.} B} {\bf 156}, 55 (1985).

\bibitem{DG}
E. Dudas and C. Grojean, {{\em Nucl. Phys.} B} {\bf 507}, 553 (1997), hep-th/9704177.

\bibitem{IQ}
I. Antoniadis and M. Quir\'{o}s,  {{\em Nucl. Phys.} B} {\bf 505}, 109 (1997),
 hep-th/9705037; I. Antoniadis and M. Quir\'{o}s, {{\em Phys. Lett.} B} {\bf 416}, 327 (1998), hep-th/9707208.

\bibitem{SS} 
J. Scherk and J. H. Schwarz,  {{\em Phys. Lett.} B} {\bf 82}, 60 (1979).

\bibitem{BT}
P. Brax and N. Turok,  {{\em Phys. Lett.} B} {\bf 413}, 331 (1997), hep-th/9706035.

\bibitem{DNNS}
M. Dine, A. E. Nelson, and Y. Shirman,  {{\em Phys. Rev.}
D} {\bf 51}, 1362 (1995), 
hep-ph/9408384;  M. Dine, A. E. Nelson, Y. Nir, and Y. Shirman, 
  {{\em Phys. Rev.} D} {\bf 53}, 2658 (1996), 
hep-ph/9507378.

\bibitem{Martin}
S. Martin,  {{\em Phys. Rev.} D} {\bf 55}, 3177 (1997), hep-ph/9608224.

\bibitem{PTr}
 E. Poppitz and S. P. Trivedi,  {{\em Phys. Lett.} B} {\bf 401}, 38 (1997), hep-ph/9703246.

\bibitem{vdBV}
J.  van der Bij and M. Veltman,  {{\em Nucl. Phys.} B} {\bf 231}, 205 (1984).

\bibitem{GRat} 
G. F. Giudice and R. Rattazzi, {{\em Nucl. Phys.} B} {\bf 511}, 25 (1998), hep-ph/9706540.

\bibitem{DGP}
S. Dimopoulos, G. F. Giudice, and A. Pomerol, {{\em Phys. Lett.}
B}{\bf 389}, 37 (1996),
hep-ph/9607225.

\bibitem{basicsusy} P. West, {\em Introduction to Supersymmetry and 
Supergravity}.  (World Scientific, Singapore, 1986).

\bibitem{wessbagger} J. Wess and J. Bagger, {\em Supersymmetry and
 Supergravity}. (Princeton University Press, Princeton, 1992).

\end{thebibliography}
\end{document}